\documentclass[aps,pra,superscriptaddress,floatfix,showpacs,10pt, twocolumn]{revtex4}
\usepackage[dvips]{graphicx}
\usepackage{amsmath,times}

\begin{document}

\title{Generation of 3-Dimensional graph state with Josephson charge qubits }
\author{Xiao-Hu Zheng\footnote{Electronic address:
xhzheng@ahu.edu.cn} and Zhuo-Liang Cao\footnote{Electronic address:
zlcao@ahu.edu.cn (Corresponding Author)}}

\affiliation{Key Laboratory of Opto-electronic Information
Acquisition and Manipulation, Ministry of Education, School of
Physics and Material Science, Anhui University, Hefei, 230039, P R
China}

\begin{abstract}
On the basis of generations of 1-dimensional and 2-dimensional graph
states, we generate a 3-dimensional $N^{3}-qubit$ graph state based
on the Josephson charge qubits. Since any two charge qubits can be
selectively and effectively coupled by a common inductance, the
controlled phase transform between any two-qubit can be performed.
Accordingly, we can generate arbitrary multi-qubit graph states
corresponding to arbitrary shape graph, which meet the expectations
of various quantum information processing schemes. All the devices
in the scheme are well within the current technology. It is a
simple, scalable and feasible scheme for the generation of various
graph states based on the Josephson charge qubits.
\end{abstract}

\maketitle

Entanglement \cite{en} can serve as basic ingredient in the course
of quantum information processing. In achieving the task of quantum
communication, the entanglement is a medium for transferring quantum
information. Owing to entanglement, quantum computers have
potentially superior computing power over their classical
counterparts. Graph states \cite{W,Hein,D} are a family of
multi-qubit states. Many well-known states, such as
Greenberger-Horne-Zeilinger (GHZ) \cite{Greenberger} states and
cluster states \cite{Briegel1}, can be generated from the graph
states. In quantum error correcting codes \cite{clark, Steane} and
in one-way quantum computing \cite{one-way1, one-way2} , some of
graph states are used as the resources. For every non-trivial graph
state it is possible to construct three-setting Bell inequalities
which are maximally violated only by this state
\cite{bellineqalities1,bellineqalities2}.

The concept of a graph is the basis of a graph state. A graph $G=(
V,E)$ comprise two classes of elements, \emph{i.e.}, vertices $V$
and edges $E$. Each graph can be represented by a diagram in a
plane, where each vertex is represented by a point and each edge $E$
by an arc joining two not necessarily distinct vertices \cite{Hein}.
For the graph states, vertices $V$ correspond to qubits of physical
systems and edges represent interactions of qubits. The state vector
$|\Psi\rangle=|+\rangle^{\otimes
V}=((|0\rangle+|1\rangle)/\sqrt{2})^{\otimes V}$  is referred to as
the graph state vector of the empty graph. The state vector of the
graph state containing edges is described as
\begin{eqnarray}
\label{1} |G\rangle &=&\prod_{(i,j\in E)}U^{(i,j)}|\Psi\rangle
\nonumber\\
&=&\prod_{(i,j\in
E)}U^{(i,j)}((|0\rangle+|1\rangle)/\sqrt{2})^{\otimes V}
\end{eqnarray}
with $U^{(i,j)} = (I^{(i)}\otimes I^{(j)}+ \sigma_z^{(i)}\otimes
I^{(j)} + I^{(i)}\otimes\sigma_z^{(j)}- \sigma_z^{(i)} \otimes
\sigma_z^{(j)})/2$ corresponding to a controlled phase-gate between
qubits labeled $i$ and $j$, described by Pauli matrices. As
mentioned above, the graph states have the special characteristics
and practical applications, so the preparation of the graph states
has become the focus of research. Clark \emph{et al.} \cite{clark}
present a scheme that allows arbitrary graph states to be
efficiently created in a linear quantum register via an auxiliary
entangling bus. Benjamin \emph{et al.} \cite{Benjamin} present a
scheme that creates graph states by simple three-level systems in
separate cavities. Bodiya \emph{et al.} \cite{Bodiya} propose a
scheme for efficient construction of graph states using realistic
linear optics, imperfect photon source and single-photon detectors.

Recently, much attention has been attracted to the quantum computer,
which works on the fundamental quantum mechanical principle. The
quantum computers can solve some problems exponentially faster than
the classical computers. For realizing quantum computing, some
physical systems, such as nuclear magnetic resonance \cite{18},
trapped irons \cite{19}, cavity quantum electrodynamics (QED)
\cite{20}, and optical systems \cite{Turchette} have been proposed.
These systems have the advantage of high quantum coherence, but
can't be integrated easily to form large-scale circuits. Because of
large-scale integration and relatively high quantum coherence,
Josephson charge qubit \cite{21-1,21-2, 22} and flux qubit
\cite{23Mooij, 24}, which are based on the macroscopic quantum
effects in low-capacitance Josephson junction circuits
\cite{Makhlin,You2005}, are the promising candidates for quantum
computing. As is well known, the graph states are mainly applied to
quantum computing. Accordingly, generation of the graph states by
Josephson charge and flux qubit is of great importance. In this
paper, we propose a scheme for the generation of the graph states
using Josephson charge qubit. This scheme is simple and easily
manipulated, because any two charge qubits can be selectively and
effectively coupled by a common inductance. More manipulations can
be realized before decoherence sets in. All of the devices in the
scheme are well within the current technology. It is a simple,
scalable and feasible scheme for the generation of various graph
states based on the Josephson charge qubits.

The paper is organized as follows: Firstly, we introduce Josephson
charge-qubit structure and Hamiltonian of the system. Secondly,
explain how to implement the controlled phase-gate. Thirdly,
illustrate the generation of the arbitrary multi-qubit graph states
corresponding to arbitrary shape graph. Fourthly, give necessary
discussions for the feasibility of our scheme. Finally, the
conclusions are given.

\begin{figure}[tbp]
\includegraphics[scale=0.8,angle=0]{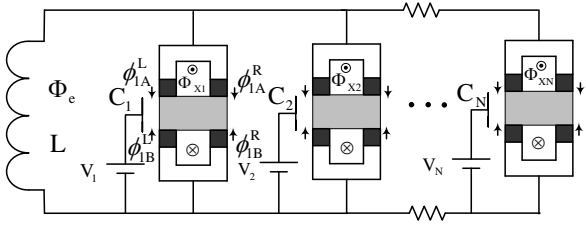}\caption{ Josephson
charge-qubit structure. Each CBP is configure both in the charging
regime $E_{ck}\gg E^0_{Jk}$ and at low temperatures $k_BT\ll
E_{ck}$. Furthermore, the superconducting gap $\Delta$ is larger
than $E_{ck}$ so that quasiparticle tunneling is suppressed in the
system. } \label{fig1}
\end{figure}

Since the earliest Josephson charge qubit scheme \cite{21-1} was
proposed, a series of  improved schemes \cite{21-2,You2002} have
been explored.  Here, we concern the  architecture of Josephson
charge qubit in Ref. \cite{You2002}, which is the first efficient
scalable quantum computing (QC) architecture. The Josephson charge
qubits structure  is shown in Fig.(\ref{fig1}). It consists of
\emph{N} cooper-pair boxes (CPBs) coupled by a common
superconducting inductance L. For the \emph{k}th  cooper-pair box, a
superconducting island with charge $Q_{k}=2en_{k}$ is weakly coupled
by two symmetric direct current superconducting quantum interference
devices (dc SQUIDs) and biased by an applied voltage through a gate
capacitance $C_{k}$. Assume that the two symmetric dc SQUIDs are
identical and all Josephson junctions in them have Josephson
coupling energy $E_{Jk}^{0}$ and capacitance $C_{Jk}$. The
self-inductance effects of each SQUID loop is usually neglected
because of the very small size ($~1\mu m$) of the loop. Each SQUID
pierced by a magnetic flux $\Phi_{Xk}$ provides an effective
coupling energy $-E_{Jk}(\Phi_{Xk})\cos\phi_{kA(B)}$, with
$E_{Jk}(\Phi_{Xk})=2E_{Jk}^{0}\cos(\pi\Phi_{Xk}/\Phi_0)$, and the
flux quantum $\Phi_0=h/2e$. The effective phase drop $\phi_{kA(B)}$,
with subscript $A(B)$ labeling the SQUID above (below) the island,
equals the average value $[\phi_{kA(B)}^L+\phi_{kA(B)}^R]/2$, of the
phase drops across the two Josephson junctions in the dc SQUID, with
superscript $L(R)$ denoting the left (right) Josephson junction.

For any given cooper-pair box, say $i$, when
$\Phi_{Xk}=\frac{1}{2}\Phi_0$ and $V_{Xk}=(2n_k+1)e/c_k$ for all
boxes except $k=i$, the inductance $L$ connects only the $i$th
cooper-pair box to form a superconducting loop. In the
spin-$\frac{1}{2}$ representation, based on charge states
$|0\rangle=|n_i\rangle$ and $|1\rangle=|n_{i+1}\rangle$, the reduced
Hamiltonian of the system becomes \cite{You2002}
\begin{equation}
\label{2}
H=\varepsilon_{i}(V_{Xi})\sigma_z^{(i)}-\overline{E}_{Ji}(\Phi_{Xi},
\Phi_e, L)\sigma_x^{(i)},
\end{equation}
where $\varepsilon_{i}(V_{Xi})$ is controlled by the gate voltage
$V_{Xi}$, while the intrabit coupling $\overline{E}_{Ji}(\Phi_{Xi},
\Phi_e, L)$ depends on inductance$L$, the applied external flux
$\Phi_e$ through the common inductance and the local flux
$\Phi_{Xi}$  through the two SQUID loops of the $\emph{i}$th
cooper-pair box. By controlling $\Phi_{Xk}$ and $V_{Xk}$, the
operations of Pauli matrice $\sigma_z^{(i)}$ and $\sigma_x^{(i)}$
are achieved. Thus, any single-qubit operations are realized by
utilizing the Eq. (\ref{1}).

To manipulate many-qubit, say $i$ and $j$, we configure
$\Phi_{Xk}=\frac{1}{2}\Phi_0$ and $V_{Xk}=(2n_k+1)e/c_k$ for all
boxes except $k=i$ and $j$. In the case, the inductance $L$ is only
shared by the cooper-pair boxes $i$ and $j$ to form superconducting
loops. The Hamiltonian of the system can be reduced to
\cite{You2002, You2001}
\begin{equation}
\label{3}
H=\sum_{k=i,j}[\varepsilon_{k}(V_{Xk})\sigma_z^{(k)}-\overline{E}_{Jk}\sigma_x^{(k)}]+\Pi_{ij}\sigma_x^{(i)}\sigma_x^{(j)},
\end{equation}
where the interbit coupling $\Pi_{ij}$ depends on both the external
flux $\Phi_e$ through the inductance $L$, the local fluxes
$\Phi_{Xi}$ and $\Phi_{Xj}$ through the SQUID loops. In Eq.
(\ref{2}), if we choose $V_{Xk}=(2n_k+1)e/c_k$, the Hamiltonian of
system can be reduced to
\begin{equation}
\label{4}
H=-\overline{E}_{Ji}\sigma_x^{(i)}-\overline{E}_{Jj}\sigma_x^{(j)}+\Pi_{ij}\sigma_x^{(i)}\sigma_x^{(j)}.
\end{equation}
For the simplicity of calculation, we assume
$\overline{E}_{Ji}=\overline{E}_{Jj}=\Pi_{ij}=\frac{-\pi\hbar}{4\tau}$($\tau$
is a given period of time), which can be obtained by suitably
choosing parameters. Thus Eq.(\ref{3}) becomes
\begin{equation}
\label{5}
H=\frac{-\pi\hbar}{4\tau}(-\sigma_x^{(i)}-\sigma_x^{(j)}+\sigma_x^{(i)}\sigma_x^{(j)}).
\end{equation}

Below, we discuss problems on the basis
$\{|+\rangle=\frac{1}{\sqrt{2}}(|0\rangle+|1\rangle),
|-\rangle=\frac{1}{\sqrt{2}}(|0\rangle-|1\rangle)\}$. According to
Hamiltonian $H$ of Eq. (\ref{5}), we can obtain the following
evolutions:
\begin{subequations}
\label{6}
\begin{equation}
|++\rangle_{ij}\rightarrow e^{-i\pi t/4\tau}|++\rangle_{ij},
\end{equation}
\begin{equation}
|+-\rangle_{ij}\rightarrow e^{-i\pi t/4\tau}|+-\rangle_{ij},
\end{equation}
\begin{equation}
|-+\rangle_{ij}\rightarrow e^{-i\pi t/4\tau}|-+\rangle_{ij},
\end{equation}
\begin{equation}
|--\rangle_{ij}\rightarrow e^{i3\pi t/4\tau}|--\rangle_{ij}.
\end{equation}
\end{subequations}
If we choose $t=\tau$, which can be achieved by choosing switching
time, and perform a single-qubit operation $U=e^{i\pi /4}$, we can
obtain
\begin{subequations}
\label{7}
\begin{eqnarray}|++\rangle_{ij}\rightarrow |++\rangle_{ij},
\end{eqnarray}
\begin{equation}
|+-\rangle_{ij}\rightarrow |+-\rangle_{ij},
\end{equation}
\begin{equation}
|-+\rangle_{ij}\rightarrow |-+\rangle_{ij},
\end{equation}
\begin{equation}
|--\rangle_{ij}\rightarrow -|--\rangle_{ij}.
\end{equation}
\end{subequations}
The Eq. (\ref{7}) have actually realized the operation of a
controlled phase gate. Any two charge qubits can be selectively and
effectively coupled by a common inductance, so the controlled phase
transform between any two-qubit is performed. It is very important
for the following generation of arbitrary multi-qubit graph states
corresponding to arbitrary shape graphs.

\begin{figure}
\includegraphics[scale=0.40,angle=0]{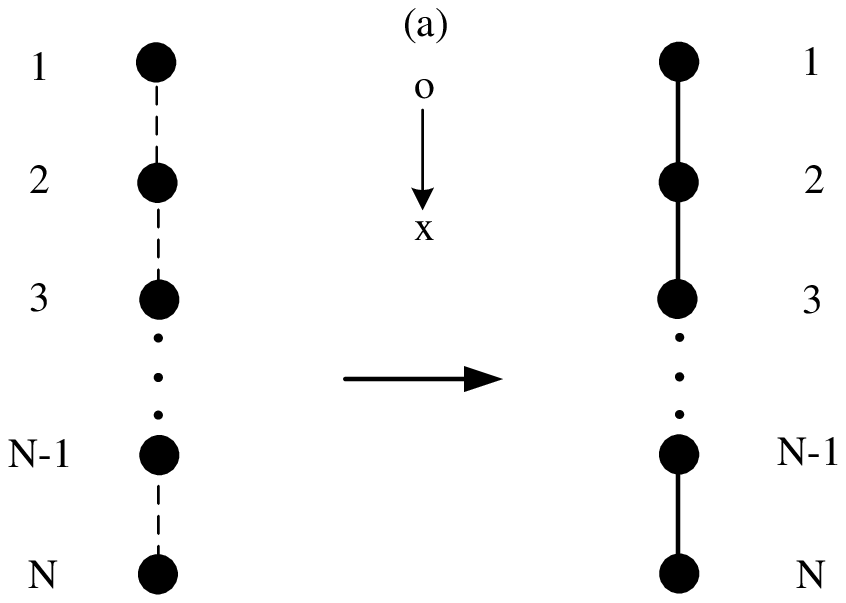}
\includegraphics[scale=0.40,angle=0]{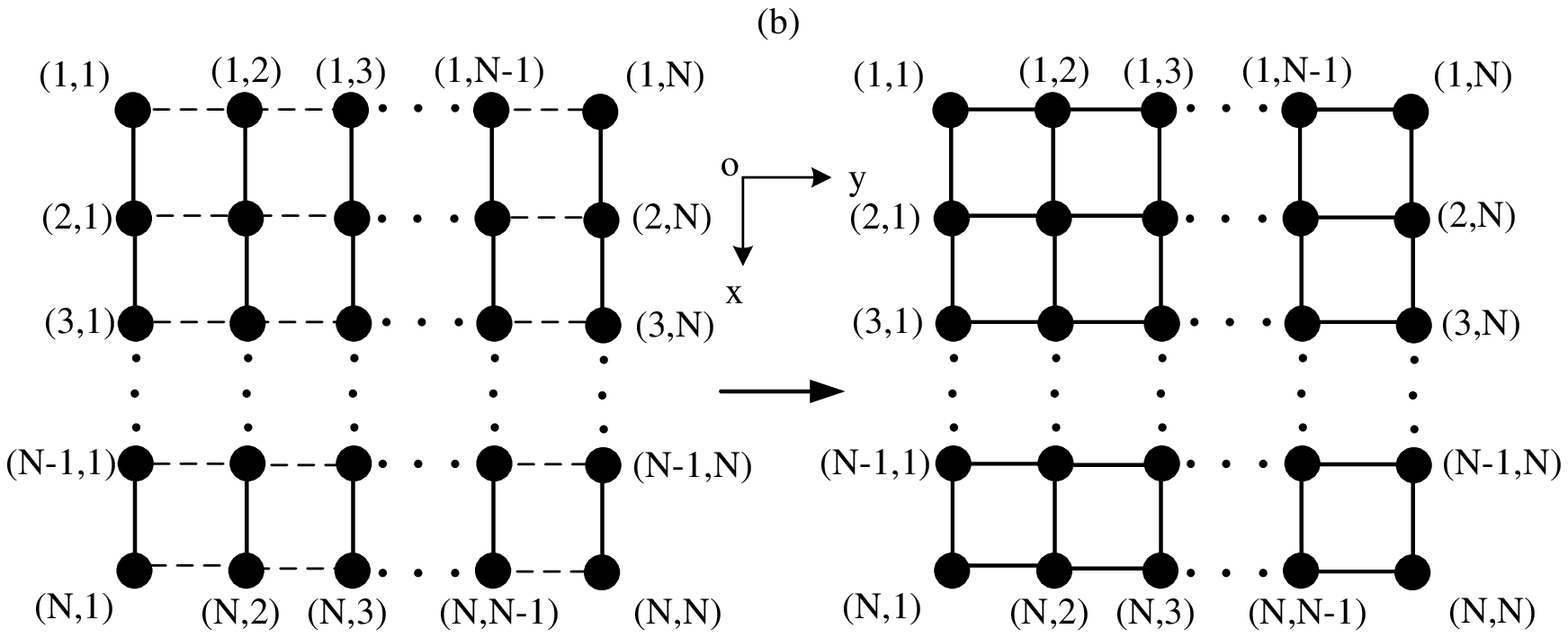}
\includegraphics[scale=0.50,angle=0]{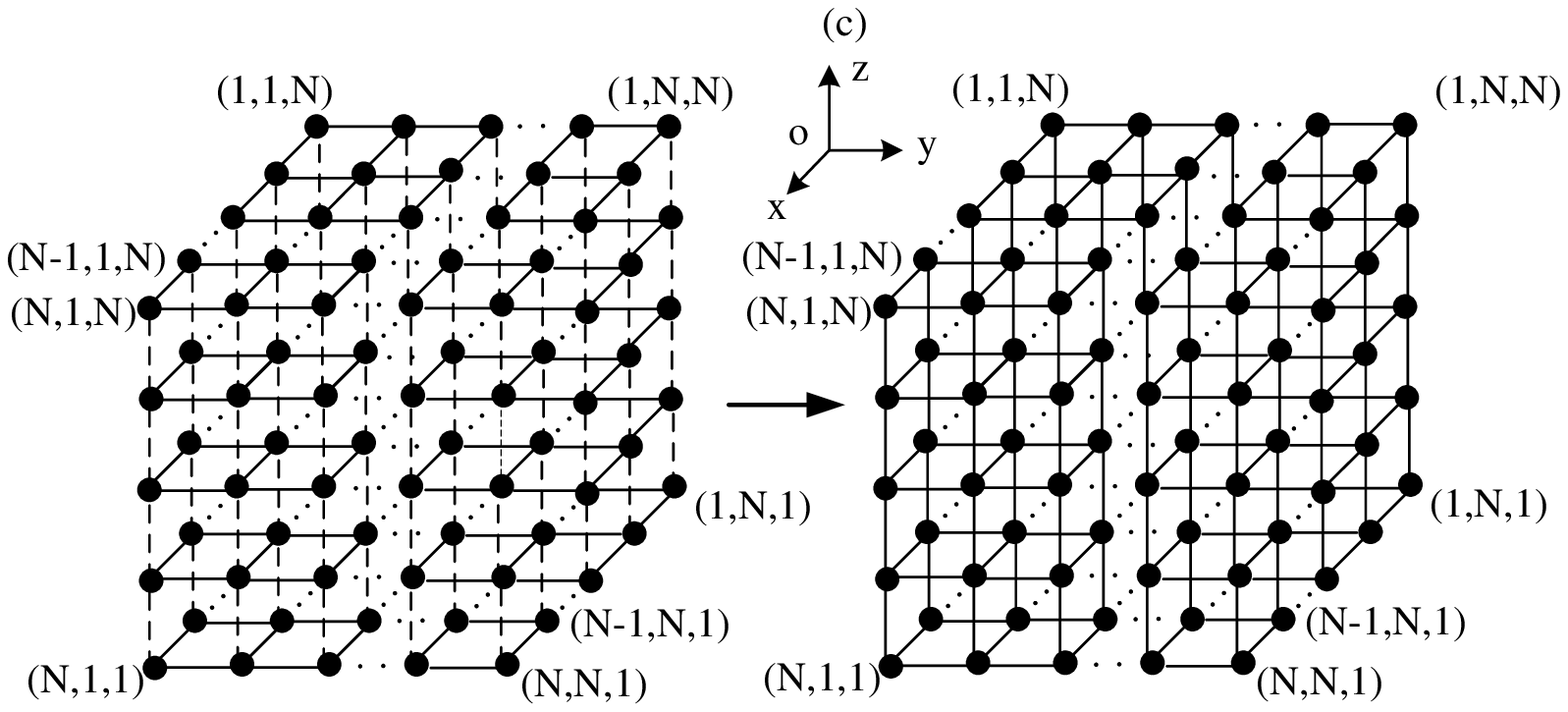}
\caption{The dashed lines in the figure denote that the interactions
between qubits haven't taken place. The real lines denote the
interactions between qubits have been completed. (a) Generation of
N-qubit graph state from the graph state of empty graph to
1-dimensional graph state. (b) Generation of the graph state from
1-dimensional graph states to 2-dimensional. (c) Generation of the
graph state from 2-dimensional graph states to 3-dimensional.}
\label{fig2}
\end{figure}

Under the basis of $\{|+\rangle, |-\rangle\}$, the state vector of
the graph state containing edges is described as
\begin{equation}
\label{8} |G\rangle=\prod_{(i,j\in
E)}U'^{(i,j)}((|+\rangle+|-\rangle)/\sqrt{2})^{\otimes V},
\end{equation}
where $U'^{(i,j)}$ is a controlled phase-gate for the basis of
$\{|+\rangle, |-\rangle\}$. Our goal is generating a 3-dimensional
$N^{3}-qubit$ graph states corresponding to a 3-dimensional graph.
The work of generating 3-dimensional graph states divides into the
following 3 steps.

Step $1$: Firstly, we prepare all N charge qubits in the states of
$|+\rangle$, which is the graph state of empty graph. Next, perform
$N-1$ controlled phase transforms between adjacent charge qubits in
$x$ axis direction as shown in fig $2(a)$. Thus we obtain a
1-dimensional graph state corresponding to the right graph of the
fig $2(a)$.

Step $2$: Firstly, we prepare N graph states of 1-dimension, which
is N-qubit graph state corresponding to the right graph in the fig
$2(a)$. Next, on the basis of the left graph of the fig $2(b)$,
perform $N(N-1)$ controlled phase transforms between adjacent charge
qubits in $y$ axis direction as shown in fig $2(b)$. Thus we obtain
a 2-dimensional graph state corresponding to the right graph in the
fig $2(b)$.

Step $3$: Firstly, we prepare N graph states of 2-dimension, which
is $N^{2}-qubit$ graph state corresponding to the right graph in the
fig $2(b)$. Next, on the basis of the left graph of the fig $2(c)$,
perform $N^2(N-1)$ controlled phase transforms between adjacent
charge qubits in $z$ axis direction as shown in fig $2(c)$. Thus we
obtain a 3-dimensional graph state, that is a $N^{3}-qubit$ graph
state corresponding to the right graph in the fig $2(c)$.

It is note to add that our scheme can generalize to generate
arbitrary multi-qubit graph states corresponding to arbitrary shape
graphs, which meet the expectations of various quantum information
processing schemes.

Below, we briefly discuss the experimental feasibility of the
current scheme. For the used charge qubit in our scheme, the typical
experimental switching time $\tau^{(1)}$ during a single-bit
operation is about $0.1ns$ \cite{You2002}. The inductance $L$ in our
used proposal is about $30nH$, which is experimentally accessible.
In the earlier design \cite{21-2}, the inductance $L$ is about
$3.6\mu H$, which is difficult to make at nanometer scales. Another
improved design \cite{Makhlin} greatly reduces the inductance $L$ to
$\sim120nH$, which is about 4 times larger than the one used in our
scheme. The fluctuations of voltage source and fluxes result in
decoherence for all charge qubits. The gate voltage fluctuation
plays the dominant role in producing decoherence. The estimated
dephasing time is $\tau_4\sim10^{-4 }s$ \cite{Makhlin}, which allow
in principle $10^6$ coherent single-bit manipulations. Owing to
using the probe junction, the phase coherence time is only about
$2ns$ \cite{Nakamura2,Nakamura2002}. In this setup, background
charge fluctuations and the probe-junction measurement may be two of
the major factors in producing decoherences \cite{You2002}. The
charge fluctuations are principal only in the low-frequency region
and can be reduced by the echo technique \cite{Nakamura2002} and by
controlling the gate voltage to the degeneracy point, but an
effective technique for suppressing charge fluctuations still needs
to be explored. According to discuss above, all of devices in our
scheme are made by the current technology.

In summary, we have investigated a simple scheme for generating the
graph states based on the Josephson charge qubit. Firstly, we
generate a 1-dimensional $N-qubit$ graph state from a graph state
corresponding to empty graph. Next, on the basis of the first step,
generate a 2-dimensional $N^{2}-qubit$ graph state. Finally, on the
basis of the second step, generate a 3-dimensional $N^{3}-qubit$
graph state. Since any two charge qubits can be selectively and
effectively coupled by a common inductance, the controlled phase
transform between any two-qubit is performed. Accordingly, we can
generate arbitrary multi-qubit graph states corresponding to
arbitrary shape graph, which meet the expectations of various
quantum information processing schemes. The architecture of our
proposal is made by present scalable microfabrication technique.
More manipulations can be realized before decoherence sets in. All
the devices in the scheme are well within the current technology. It
is a simple, scalable and feasible scheme for the generation of
various graph states based on the Josephson charge qubits.

\begin{acknowledgments}
This work is supported by Natural Science Foundation of China under
Grants No. 60678022 and No. 10674001, the Key Program of the
Education Department of Anhui Province under Grant No. 2006KJ070A,
the Program of the Education Department of Anhui Province under
Grant No. 2006KJ057B and the Talent Foundation of Anhui University.

\end{acknowledgments}

\end{document}